\documentclass[dvipdfmx,conference]{IEEEtran}
\usepackage{amssymb,stmaryrd,amsmath,amsfonts,rotating}%amsfonts,amsthm
\usepackage{color}
\usepackage{bbm}
\usepackage{latexsym}
 
\newcommand{\qed}{\hfill \ensuremath{\Box}}

\usepackage{mathtools}
\mathtoolsset{showonlyrefs}

\newtheorem{theorem}{Theorem}
\newtheorem{definition}{Definition}

\newtheorem{lemma}{Lemma}
\newtheorem{discussion}{Discussion}

\newcommand{\F}{\mathbb{F}}

\newcommand{\GF}{\mathrm{GF}}
\newcommand{\GL}{\mathrm{GL}}

\newcommand{\dl}{d_{{l}}}

\newcommand{\tell}{t}
\newcommand{\dr}{d_{{r}}}

\usepackage{mathtools}
\mathtoolsset{showonlyrefs}

\ifCLASSINFOpdf
\else
\fi
\begin{document}
\title{Non-Binary LDPC Codes with Large Alphabet Size}
\author{
\IEEEauthorblockN{Koji Tazoe,  Kenta Kasai and Kohichi Sakaniwa}
\IEEEauthorblockA{Department of Communications and Computer Engineering,\\
 Tokyo Institution of Technology\\
Email: \{tazoe,kenta,sakaniwa\}@comm.ce.titech.ac.jp}

}
\maketitle

\begin{abstract}
We study LDPC codes for  the channel with input ${x}\in \mathbb{F}_q^m$ and 
output ${y}={x}+{z}\in \mathbb{F}_q^m$. 
The aim of this paper is to evaluate decoding performance of $q^m$-ary non-binary LDPC codes for large $m$. 
We give density evolution and decoding performance evaluation for regular non-binary LDPC codes and spatially-coupled (SC) codes. 
We show the regular codes do not achieve the capacity of the channel while SC codes do. 
\end{abstract}
\IEEEpeerreviewmaketitle

%%%%%%%%%%%%%%%%%%%%%%%%%%%%%%%%%%%%%%%%%%%%%%%%%%%%%%%%%%%%%%%
\section{Introduction}
%%%%%%%%%%%%%%%%%%%%%%%%%%%%%%%%%%%%%%%%%%%%%%%%%%%%%%%%%%%%%%%
In 1963, Gallager invented low-density parity-check (LDPC) codes \cite{gallager_LDPC}.
Due to sparsity of the code representation, 
LDPC codes are efficiently decoded by belief propagation (BP) decoders. 
By a powerful optimization method {\it density evolution} \cite{910577}, developed by Richardson and Urbanke,
messages of BP decoding can be statistically evaluated. 
The optimized LDPC codes can approach very close to Shannon limit \cite{richardson01design}.

In this paper, we consider non-binary LDPC codes over $\F_q^m$ defined by sparse parity-check matrices over $\GL(m,\F_q)$
Non-binary LDPC codes were invented by Gallager \cite{gallager_LDPC}. 
Davey and MacKay \cite{DaveyMacKayGFq} found non-binary LDPC codes can outperform binary ones. 
Non-binary LDPC codes have captured much attention recently due to their decoding performance \cite{KTRS_QKD_ISITA},\cite{nb_lr_IT},\cite{6220213},\cite{6017122}. 
It is observed $2^m$-ary non-binary codes exhibit excellent decoding performance around at $m=6$ over BMS channels. 

Spatially-coupled (SC) codes attract much attention due to their capacity-achieving performance and a memory-efficient sliding-window decoding algorithm.
Recently, SC codes are shown to prove achieve capacity of BEC \cite{5706927}, \cite{ISIT_OJKP} and BMS channels \cite{2012arXiv1201.2999K}.

In this paper, we study coding over the channel with input ${x}\in \mathbb{F}_q^m$ and output ${y}\in \mathbb{F}_q^m$. 
The receiver knows a subspace $V\subset \mathbb{F}_q^m$ from which ${z}={y}-{x}$ is uniformly chosen.
Or equivalently, the receiver receives an affine subspace ${y}-V:=\{{y}-{z}\mid{z}\in V\}$ in which the input ${x}$ is compatible. 
This channel model is used in the decoding process for network coding \cite{6397614} after estimating noise packet spaces. 
In \cite{6284231},  the authors proposed a coding scheme with binary SC MacKay-Neal codes with the joint iterative decoding between the channel detector and the code decoder. 
It was observed that the code exhibits capacity achieving performance for small $m$. 
The channel detector calculates log likelihood ratio (LLR) of the transmitted bits from a channel output and messages from the BP decoder.

The aim of this paper is to evaluate decoding performance of $q^m$-ary non-binary LDPC codes for large $m$. 
We give density evolution and decoding performance evaluation for regular non-binary LDPC codes and SC codes. 
We show the regular codes do not achieve the capacity of the channel while SC codes do. 
%%%%%%%%%%%%%%%%%%%%%%%%%%%%%%%%%%%%%%%%%%
\section{Channel Model}
%%%%%%%%%%%%%%%%%%%%%%%%%%%%%%%%%%%%%%%%%%
In this paper, we consider channels with input ${x}\in \mathbb{F}_q^m$ and 
output ${y}={x}+{z}\in \mathbb{F}_q^m$, where  ${z}\in \mathbb{F}_q^m$ is 
uniformly distributed in a linear subspace $V\subset\mathbb{F}_q^m$ of dimension $\epsilon m$. 
It is easy to see that the channel is weakly symmetric \cite{cover06}. 
From \cite[Theorem 7.2.1]{cover06}, the normalized capacity is given by 
\begin{align*}
C= \frac{1}{m}\max_{p({X})}I({X};{Y}) = (1-\epsilon). 
\end{align*}
The channel with  large $m$ was used in a decoding process of the network coding scenario \cite{DBLP:journals/corr/abs-0711-3935}. 
In \cite{DBLP:journals/corr/abs-0711-3935}, the data part of each packet is represented as ${x}\in\mathbb{F}_q^m$. 
Packets are coded by non-binary LDPC codes whose parity-check coefficients are in the general linear group $\GL(m,\mathbb{F}_q)$. 
The noise subspace $V$ is estimated by padding zero packets and using Gaussian elimination. 
We denote this channel by $\mathrm{CD}(m, \epsilon)$. 
%%%%%%%%%%%%%%%%%%%%%%%%%%%%%%%%%%%%%%%%%%
\section{Code Definition}
%%%%%%%%%%%%%%%%%%%%%%%%%%%%%%%%%%%%%%%%%%
In this section, we briefly review  $(\dl,\dr)$ codes and $(\dl,\dr, L)$  codes introduced by Kudekar 
{\it et al.}~\cite{5695130}. We assume $\frac{\dr}{\dl}\in\mathbb{Z}$ and $\frac{\dr}{\dl}\ge 2$.
Both $(\dl,\dr)$ codes and $(\dl,\dr, L)$  codes are defined over $\GF(q)$ and have parity-check matrix over $\GF(q)$. 

%---------------------------------------
\subsection{$(\dl,\dr)$-Codes}
%---------------------------------------
Let $H(\dl, \dr)$ be an $M\dl\times M\dr$ sparse binary matrix of column weight $\dl$ and row weight $\dr$. 
The Tanner graph of ($\dl,\dr,L$)  code is obtained by making $M$ copies of protographs of $H(\dl,\dr,L)$ and connecting edges among the same edge types. 
$H(\dl,\dr,m)$ is given by replacing 1 with a randomly chosen non-zero elements in $\GL(m,\F_q)$ and replacing 0 with $0\in\GL(m,\F_q)$, 
where $\GL(m, \F_q)$ is the set of all non-singular $\F_q$-valued matrix  of size $m\times m$. 
The resultant matrix $H(\dl,\dr,m)$ can be viewed as a $\GL(m,\F_q)$-valued matrix of size $M\dl\times M\dr$. 
%---------------------------------------
\subsection{$(\dl,\dr,L)$-Codes}
%---------------------------------------
The $(\dl,\dr, L)$ codes are defined by the following protograph codes \cite{protograph}. 
The adjacency matrix of the protograph is referred to as a base matrix. 
The base matrix of ($\dl,\dr,L$)  code is given as follow. 
Let $H(\dl,\allowbreak \dr,\allowbreak L)$ be an $(L+\dl-1)\times \frac{\dr}{\dl} L$ band binary matrix 
of band size $\dr\times\dl$ and column weight $\dl$, where the band size is height $\times$ width of the band.
We refer to $L$ as coupling number. 
For example 
\begin{center}
 $ H(\dl=4,\dr=8,L=9)=
 \left[
 \begin{minipage}{4.0cm}
 \vspace{1mm} 
 \renewcommand{\baselinestretch}{0.3}
 \begin{verbatim}
 11                
 1111              
 111111            
 11111111          
   11111111        
     11111111      
       11111111    
         11111111  
           11111111
             111111
               1111
                 11
 \end{verbatim}
 \renewcommand{\baselinestretch}{1.0}
 \end{minipage}
 \right].
 $
\end{center}
The Tanner graph of ($\dl,\dr,L$)  code is obtained by making $M$ copies of protographs of $H(\dl,\dr,L)$ and connecting edges among the same edge types. 
The parameter $M$ is referred to as lifting number. 
The  matrix $H(\dl,\dr,L,M)$ is given by replacing each 1 in $H(\dl,\dr,L)$ with an $M\times M$ random permutation  and 
each 0 with an $M\times M$ zero matrix.
%Let $P_{i,j}$ for $i \in [1,L+\dl-1]$, $j \in [1,\dr]$ be a binary random $M \times M$ permutation matrices. 
$H(\dl,\dr,L,M,m)$ is given by replacing 1 with a randomly chosen non-zero elements in $\GL(m,\F_q)$ and replacing 0 with $0\in\GL(m,\F_q)$, 
where $\GL(m, \F_q)$ is the set of all non-singular $\F_q$-valued matrix  of size $m\times m$. 
The resultant matrix $H(\dl,\dr,L,M,m)$ can be viewed as a $\GL(m,\F_q)$-valued matrix of size $(L+\dl-1)M \times \frac{\dr}{\dl} LM$. 
%%%%%%%%%%%%%%%%%%%%%%%%%%%%%%%%%%%%%%%%%%%%%%%
\section{Decoding Algorithm}
%%%%%%%%%%%%%%%%%%%%%%%%%%%%%%%%%%%%%%%%%%%%%%%
Let $\mathcal{H}$ be a $\GL(m,\F_q)$-valued matrix given by the construction above. 
Denote row  and column size of $\mathcal{H}$ by $\mathcal{M}$ and $\mathcal{N}$, respectively. 
Denote the $(i,j)$-th entry of $\mathcal{H}$ by $h_{i,j}\in \GL(m,\F_q)$. Then a codeword $(x_1,\dotsc,x_{\mathcal{N}})\in (\F_q^m)^{\mathcal{N}}$ satisfies parity-check equations
\begin{align}
 \sum_{j\in \partial i}h_{i,j}{x}_{j}=0, 
\end{align}
for $i=1,\dotsc, \mathcal{M}$
where $\partial i:=\{j\in \{1,\dotsc,\mathcal{N}\}\mid h_{i,j}\neq 0\}$. 

Sum-product algorithm (SPA) \cite{910572} is employed to decode. 
Without loss of generality, we can assume all-zero codeword was sent to make analysis easier \cite{4111/THESES}. 
The SPA tries to marginalize the following function with respect to each $x_j\ (j=1,\dotsc, \mathcal{N}). $
\begin{align}
\prod_{j=1}^{\mathcal{N}}\Pr({Y}_{j}={y}_j\mid {X}_j={x}_j)
\prod_{i=1}^{\mathcal{M}}\mathbbm{1}\Bigl[\sum_{j\in \partial i}h_{j,i}{x}_{j}=0\Bigr],
\end{align}
where $\mathbbm{1}[\cdot]$ is the indicator function.
The SPA message forms a uniform probability vector over a subset of $\F_q^m$. 
The support of each sum-product message forms a linear subspace of $\F_q^m$ \cite{4111/THESES}. 

%%%%%%%%%%%%%%%%%%%%%%%%%%%%%%%%%%%%%%%%%%%%%%%%%%%%%%%%%%%%%%%%%%%%%%%%%%%%%%%%%%%%%%%%%%%%%%
\section{Density Evolution Analysis of $(\dl,\dr)$-Codes}
%%%%%%%%%%%%%%%%%%%%%%%%%%%%%%%%%%%%%%%%%%%%%%%%%%%%%%%%%%%%%%%%%%%%%%%%%%%%%%%%%%%%%%%%%%%%%%
Denote the message subspace sent along a randomly picked edge connecting symbol nodes 
to check nodes  at the $\tell$-th iteration by $V^{(\tell)}$. 
Similarly, denote the message subspace sent along a randomly picked edge connecting check nodes 
to symbol nodes  at the $\tell$-th iteration by $U^{(\tell)}$.  
The initial message subspace $V^{(0)}$ is given by a uniformly random subspace of dimension $m\epsilon$. 
Density evolution \cite{4111/THESES} gives update equations of $V^{(\tell)}$ and $U^{(\tell)}$ as follows. 
\begin{align}
 U^{(\tell)}&=\sum_{i=1}^{\dr-1}V_i^{(\tell)},  \\
V^{(\tell)}&=V^{(0)}\cap\bigcap_{i=1}^{\dl-1}U_i^{(\tell)}.
\end{align}
where $U_i^{(\tell)}$ and $V_i^{(\tell)}$ are iid copies of $U^{(\tell)}$ and $V^{(\tell)}$, respectively and $V_1+V_2:=\{v_1+v_2\mid v_1\in V_1, v_2\in V_2\}$. 
If $V^{(\tell)}$ becomes $\{0\}$, decoding is successfully completed. 

It is not easy to track $V^{(\tell)}$. 
Instead, we track the dimension of $V^{(\tell)}$. 
We define $\xi^{(\tell)}$ in order to predict the $\dim V^{(\tell)}$. 
\begin{definition}
 Define 
 \begin{align}
   \zeta^{(\tell+1)}&= (\xi^{(\tell)})^{\boxplus(\dr-1)} \label{DE1}\\
     \xi^{(\tell)}  &=\epsilon\boxdot (\zeta^{(\tell)})^{\boxdot(\dl-1)}, \label{DE2} \\
 \xi^{(0)}  &= \epsilon
 \end{align}
where for $\xi_1, \xi_2\in [0,1]$
 \begin{align}
 \xi_1\boxdot \xi_2:&=\max(\xi_1+\xi_2 - 1,0),\\
 \xi_1\boxplus \xi_2:&=\min(\xi_1+\xi_2,1).
 \end{align}
\end{definition}

Next Lemma shows $\frac{1}{m}\dim V^{(\tell)}$ converges to $\xi^{(\tell)}$ in probability. 
\begin{lemma}
\label{lemma1}
For any $\delta>0$ and $\epsilon>0$, there exists $m'$ such that for $m>m'$
\begin{align}
  &\Pr\{ |\dim V^{(\tell)}-\xi^{(\tell)}m|<\delta m\} > 1-\epsilon. 
\end{align}
\end{lemma}
{\itshape Proof}: 
Let $V_1$ be a uniformly random subspace of dimension $d_1$ in $\mathbb{F}_q^{m}$,
and $V_2$ a uniformly random subspace  of dimension $d_2$.
Then from \cite[Proposition 4.4]{6397614}, it holds that for any $k\ge 0$ and $m\ge 0$, 
\begin{align}
 &\Pr\{ {d_1} \boxdot {d_2}\le {\dim(V_1 \cap V_2)} <  {d_1} \boxdot {d_2} +k \} \\
 &\qquad \ge 1-q^{-k-\max(0,m-d_1-d_2)} \label{eq1}, \\
 &\Pr\{ {d_1} \boxplus {d_2}-k\le {\dim(V_1 + V_2)} <  {d_1} \boxplus {d_2}  \} \\
 &\qquad \ge 1-q^{-k-\max(0,m-d_1-d_2)}, \label{eq2}
\end{align}
where,  with abuse of notation, we define $\boxdot$ and $\boxplus$ for $d_1, d_2\in \mathbb{N}$  as follows
\begin{align}
 d_1\boxdot d_2:&=\max(d_1+d_2 - m,0),\\
 d_1\boxplus d_2:&=\min(d_1+d_2,m).
\end{align}
For $\xi_1:=d_1/m$ and $\xi_2:=d_2/m$ it follows that
\begin{align}
&\Pr\Bigl\{ \Big|\frac{\dim(V_1 \cap V_2)}{m} -{\xi_1} \boxdot {\xi_2}\Big|<   \frac{k}{m} \Bigr\} \\
%&=\Pr\{ |{\dim(V_1 \cap V_2)} -{d_1} \boxdot {d_2}|<   k \} \\
&\ge\Pr\{ {d_1} \boxdot {d_2}\le {\dim(V_1 \cap V_2)} <  {d_1} \boxdot {d_2} +k \}\\
%&\ge 1-q^{-k-\max(0,m-d_1-d_2)}\\
&\ge 1-q^{-k-m\max(0,1-\xi_1-\xi_2)}. 
\end{align}
From this, for sufficiently large $m$ such that $\frac{k}{m}<\delta$ and $q^{-k-m\max(0,1-\xi_1-\xi_2)}<\epsilon$, it holds that 
\begin{align}
&\Pr\Bigl\{ \Big|\frac{\dim(V_1 \cap V_2)}{m} -{\xi_1} \boxdot {\xi_2}\Big|<   \delta \Bigr\} \ge 1-\epsilon. 
\end{align}
Similarly, we have 
\begin{align}
&\Pr\Bigl\{ \Big|\frac{\dim((V_1 \cap V_2)\cap V_3)}{m} -\frac{\dim(V_1 \cap V_2)}{m} \boxdot {\xi_3}\Big|<   \delta \Bigr\} \\
&\ge 1-\epsilon. 
\end{align}
The union bound of the two probabilities gives
\begin{align}
\Pr\Bigl\{ &\Big|\frac{\dim(V_1 \cap V_2)}{m} -{\xi_1} \boxdot {\xi_2}\Big|<   \delta \mbox{ and }\\
 &\Big|\frac{\dim((V_1 \cap V_2)\cap V_3)}{m} -\frac{\dim(V_1 \cap V_2)}{m} \boxdot {\xi_3}\Big|<   \delta \Bigr\}\\
&\ge 1- 2\epsilon
\end{align}
Using the triangle inequality and the fact that $\boxdot$ is a continuous function, we have
\begin{align}
&\Pr\Bigl\{ \Big|\frac{\dim((V_1 \cap V_2)\cap V_3)}{m} - ({\xi_1} \boxdot {\xi_2} )\boxdot {\xi_3}\Big|<   2\delta \Bigr\}\\
&\ge 1- 2\epsilon. 
\end{align}
The same argument is valid for any combinations of $\boxdot$ and $\boxplus$ of $V_i^{(0)}$ ($i=0,1,\dotsc$). $V^{(\tell)}$ is an instance of the combinations. 
Hence the thesis holds. 
\begin{align}
  &\Pr\{ |\dim V^{(\tell)}-\xi^{(\tell)}m|<\delta m\} > 1-\epsilon. 
\end{align}
\begin{discussion}
 From Lemma \ref{lemma1}, it follows that even a single parity-check code is enough to achieve the capacity when $m$ is infinite. 
However the aim of this paper is not  to design codes for $\mathrm{CD}(m,\epsilon)$, but evaluate the performance of non-binary codes for large $m$. 
\end{discussion}
\begin{lemma}
\label{lemma2}
\begin{align}
 &  \sup\{\epsilon\in [0,1]\mid \lim_{\tell\to\infty}\xi^{(\tell)}=0\}=\frac{1}{\dr-1}.
\end{align}
\end{lemma}
{\itshape Proof}:
 It is easy to see that 
\begin{align}
 \xi^{\boxplus(\dr-1)}&=\min((\dr-1)\xi,1),\\
 \epsilon \boxdot\xi^{\boxdot(\dl-1)}&=\max((\dl-1)\xi+\epsilon-(\dl-1),0). 
\end{align}
First, we claim that $ \xi^{(\tell)} \ge \frac{1}{\dr-1}$ for $\tell\ge 1$ if $\epsilon \ge \frac{1}{\dr-1}$. We use induction. Under the assumption that $\xi^{(\tell)} \ge \frac{1}{\dr-1}$, 
we can see that 
\begin{align}
\zeta^{(\tell+1)} &= \min((\dr-1)\xi^{(\tell)},1) = 1\\
  \xi^{(\tell+1)} &= \max((\dl-1)\zeta^{(\tell+1)}+\epsilon-(\dl-1),0) = \epsilon. 
\end{align}
Hence, we obtain that for all $\tell\ge 0$, 
\begin{align}
  \xi^{(\tell)} = \xi^{(0)} \ge \frac{1}{\dr-1}.
\end{align}
Next, we claim that $\lim_{\tell \to \infty} \xi^{(\tell)}=0$  if $0 \le \epsilon < \frac{1}{\dr-1}$.
It follows that  $0 \le \zeta^{(\tell)} < 1$, \eqref{DE1} and \eqref{DE2} can be rewritten respectively by 
\begin{align}
  \xi^{(\tell+1)} =& \max\Bigl((\dl-1)(\dr-1)\xi^{(\tell)} + \epsilon - (\dl -1),0\Bigr). 
\end{align}
This can be solved as 
\begin{align}
  \xi^{(\tell)} = &\max\biggl(\frac{(\dl-1)\{(\dl-1)(\dr-1)\}^\tell}{(\dl-1)(\dr-1)-1}((\dr-1)\epsilon -1)\\
&+ \frac{\epsilon-(\dl-1)}{1-(\dl-1)(\dr-1)},0\biggr)
\end{align}
From this, it can be seen that if $\epsilon < \frac{1}{\dr-1}$, $\xi^{(\tell)}$ is monotonically decreasing down to 0. \qed

We define the threshold which shows how good the ($\dl,\dr$) code is. 
For $\epsilon<\epsilon(\dl,\dr)$, ($\dl,\dr$) codes achieve vanishing decoding error probability. 
\begin{definition}
We define the \it{threshold} of ($\dl,\dr$) codes as follows.  
\begin{align}
\epsilon(\dl,\dr)=   \sup\{\epsilon\in [0,1]\mid \lim_{\tell\to\infty}\lim_{m\to\infty}\dim V^{(\tell)}=0\}.
\end{align}
We say that the ($\dl,\dr$) codes achieve capacity of $\mathrm{CD}(m, \epsilon)$ when $\epsilon(\dl,\dr)=\frac{\dl}{\dr}$. 
\end{definition}

From Lemma \ref{lemma1},  Lemma \ref{lemma2} we have the following theorem. 
\begin{theorem}
For $\dl\ge 2$,   $\epsilon(\dl,\dr)=\frac{1}{\dr-1}$. 
\end{theorem}
%--------
%%%%%%%%%%%%%%%%%%%%%%%%%%%%%%%%%%%%%%%%%%%%%%%%%%%%%%%%%%%%%%%%%%%%%%%%%%%%%%%%%%%%%%%%%%%%%%
\section{Density Evolution Analysis of $(\dl,\dr,L)$-Codes}
%%%%%%%%%%%%%%%%%%%%%%%%%%%%%%%%%%%%%%%%%%%%%%%%%%%%%%%%%%%%%%%%%%%%%%%%%%%%%%%%%%%%%%%%%%%%%%
Denote the message subspace sent along a randomly picked edge connecting symbol nodes 
to check nodes  at the $\tell$-th iteration from section $i$ to section $j$ by $V_{i,j}^{(\tell)}$. 
Similarly, denote the message subspace sent along a randomly picked edge connecting check nodes 
to symbol nodes  at the $\tell$-th iteration  from section $i$ to section $j$ by $U_{i,j}^{(\tell)}$.  

The initial message subspace $V_i^{(0)}$ is given by a uniformly random subspace of dimension $m\epsilon$ for $i\in\{0,\dotsc,L-1\}$ and $V_i^{(0)}=\{0\}$ for $i\notin\{0,\dotsc,L-1\}$. 
Density evolution gives update equations of $V^{(\tell)}$ and $U^{(\tell)}$ as follows. 
\begin{align}
V^{(0)}_{i,i}  &=V^{(0)}_{i,i+1}=\cdots=V^{(0)}_{i,i+\dl-1}=V_i^{(0)},\\
U^{(\tell+1)}_{i,j}&=\sum_{k=0,k\neq j}^{\dl-1} V^{(\tell)}_{i-k,k},\\
V^{(\tell)}_{i,j}  &=V_i^{(0)}\cap\Bigl(\bigcap_{k=0,k\neq j}^{\dl-1} U^{(\tell)}_{i+k,k}\Bigr),\\ 
V^{(\tell)}_{i}  &=V_i^{(0)}\cap\Bigl(\bigcap_{k=0}^{\dl-1} U^{(\tell)}_{i+k,k}\Bigr). \label{165654_1Oct13}
\end{align}

\begin{definition}
For $i\notin\{0,\dotsc,L-1\}$, we set define
\begin{align}
 \xi^{(0)}_i&=\xi^{(0)}_{i,j}=0
\end{align}
For $i\in\{0,\dotsc,L-1\}$, define
\begin{align}
\xi^{(0)}_{i}&=\xi^{(0)}_{i,j}=\epsilon,\\
\zeta^{(\tell+1)}_{i,j}&=\boxplus_{k=0,k\neq j}^{\dl-1} \xi^{(\tell)}_{i-k,k},\\
\xi^{(\tell)}_{i,j}  &=\epsilon\boxdot\bigl(\boxdot_{k=0,k\neq j}^{\dl-1} \zeta^{(\tell)}_{i+k,k}\bigr),\\
\xi^{(\tell)}_{i}  &=\epsilon\boxdot\bigl(\boxdot_{k=0}^{\dl-1} \zeta^{(\tell)}_{i+k,k}\bigr).
\end{align}
\end{definition}
%---------------------------
\begin{lemma}
\label{lemma3}
For any $\delta>0$ and $\epsilon>0$, there exists $m'$ such that for $m>m'$
\begin{align}
  &\Pr\{ |\dim V_{i,j}^{(\tell)}-\xi_{i,j}^{(\tell)}m|<\delta m\} > 1-\epsilon, \\
  &\Pr\{ |\dim V_{i}^{(\tell)}-\xi_{i}^{(\tell)}m|<\delta m\} > 1-\epsilon. 
\end{align}
\end{lemma}
{\itshape Proof}: 
The proof is similar to that of Lemma \ref{lemma1} and hence omitted. \qed
%--------
\begin{lemma}
\label{lemma4}
\begin{align}
 &  \sup\Bigl\{\epsilon\in [0,1]\mid \lim_{\tell\to\infty}\xi_i^{(\tell)}=0, \ i=0,\dotsc,L-1\Bigr\}=\frac{\dl}{\dr}. 
\end{align}
\end{lemma}
{\itshape Proof}: It sufficient to show  that if $\epsilon=\frac{\dl}{\dr}$,   $\xi_i=0$. This is due to the fact that  $\frac{\dl}{\dr}$ is the Shannon threshold. 
First let us check messages from check nodes at section 0 to variable nodes at section 0. 
\begin{align}
 \zeta^{(1)}_{0,0}&=\overbrace{\epsilon\boxplus \cdots \boxplus \epsilon}^{\dr-1}=\frac{\dl}{\dr}\Bigl(\frac{\dr}{\dl}-1\Bigr)=1-\frac{\dl}{\dr}. 
\end{align}
We employ peeling decoder \cite[p. 30]{4111/THESES} instead of SPA at section 0. The threshold should be the same \cite{4111/THESES}. 
\begin{align}
 \xi^{(1)}_{0}&= \zeta^{(1)}_{0,0}+\zeta^{(1)}_{0,1}+\cdots+\zeta^{(1)}_{0,\dl-1}+\epsilon-\dl\\
 &\le \zeta^{(1)}_{0,0}+1+\cdots+1+\epsilon-\dl\\
 &= \zeta^{(1)}_{0,0}+\epsilon-1=0. 
\end{align}
This implies all symbols at section 0 can be successfully decoded. 
This reduces $(\dl,\dr,L)$-code to  $(\dl,\dr,L-1)$-code. 
Repeat the decoding step $L$ times then all symbols will be decoded. \qed

\begin{definition}
We define BP threshold of ($\dl,\dr,L$) codes as follows.  
\begin{align}
\epsilon(\dl,\dr,L)=   \sup\{\epsilon\in [0,1]\mid \lim_{\tell\to\infty}\lim_{m\to\infty}\dim V_i^{(\tell)}=0\}, 
\end{align}
 where $V_i^{(\tell)}$ is defined in \eqref{165654_1Oct13}. 
\end{definition}
From Lemma \ref{lemma3},  Lemma \ref{lemma4} and the fact that the $(\dl,\dr,L)$  codes have rate $1-\frac{\dl}{\dl}-\frac{\dl-1}{L}$, 
we have the following theorem. 
\begin{theorem}
 In the limit of large $m$, the  $(\dl,\dr,L)$  codes have threshold $1-\frac{\dl}{\dr}$. 
 In the limit of large coupling number $L$, the  $(\dl,\dr,L)$  codes achieve the capacity of $\mathrm{CD}(m, \epsilon)$. 
\begin{align}
 &\lim_{L\to\infty}\epsilon(\dl,\dr,L)=\frac{\dl}{\dr}, \\
 &\lim_{L\to\infty}\lim_{m\to\infty}R(\dl,\dr,L)=1-\frac{\dl}{\dr}
\end{align}
\end{theorem}

\section{Conclusion}
We investigated decoding performance of $q^m$-ary non-binary LDPC codes for large $m$ over $\mathrm{CD}(m, \epsilon)$. 
We gave density evolution and decoding performance evaluation for regular non-binary LDPC codes and SC codes. 
We show the regular codes do not achieve the capacity of the channel while SC codes do. 
%%%%%%%%%%%%%%%%%%%%%%%%%%%%%%%%%%%%%%%%%%%%%%%%%%%%%
 \section{Conclusion}
%%%%%%%%%%%%%%%%%%%%%%%%%%%%%%%%%%%%%%%%%%%%%%%%%%%%%
\bibliographystyle{IEEEtran}
\bibliography{IEEEabrv,../../kenta_bib}
\end{document}